\documentclass[aip,jcp,reprint]{revtex4-1}

\usepackage{color}
\usepackage{dcolumn}
\usepackage{amsmath}
\usepackage{amssymb}
\usepackage{mathtools}
\usepackage{graphicx}
\usepackage{latexsym}
\usepackage{amsfonts}
\usepackage{amssymb}
\usepackage{comment}
\DeclareGraphicsExtensions{.pdf,.gif,.jpg}

\newcommand{\be}{\begin{equation}}
\newcommand{\ee}{\end{equation}}
\newcommand{\beq}{\begin{eqnarray}}
\newcommand{\eeq}{\end{eqnarray}}

\def\partl{\stackrel{\leftarrow}{\partial}}
\def\tb{\bar{t}}
\def\a{\alpha}
\def\b{\beta}
\def\g{\gamma}

\def\im{{\mathrm{i}}}
\def\ex{{\mathrm{e}}}
\def\ud{\mathrm{d}}

\def\br{\mbox{\boldmath $r$}}
\def\brp{\mbox{\boldmath $r'$}}

\draft 

\begin{document}


\title{Efficient computation of the second-Born self-energy using tensor-contraction operations} 



\author{Riku Tuovinen}
\email[]{riku.tuovinen@mpsd.mpg.de}
\affiliation{Max Planck Institute for the Structure and Dynamics of Matter, \\Luruper Chaussee 149, 22761 Hamburg, Germany}
\author{Fabio Covito}
\affiliation{Max Planck Institute for the Structure and Dynamics of Matter, \\Luruper Chaussee 149, 22761 Hamburg, Germany}
\author{Michael A. Sentef}
\affiliation{Max Planck Institute for the Structure and Dynamics of Matter, \\Luruper Chaussee 149, 22761 Hamburg, Germany}


\date{\today}

\begin{abstract}
In the nonequilibrium Green's function approach, the approximation of the correlation self-energy at the second-Born level is of particular interest, since it allows for a maximal speed-up in computational scaling when used together with the Generalized Kadanoff-Baym Ansatz for the Green's function. The present day numerical time-propagation algorithms for the Green's function are able to tackle first principles simulations of atoms and molecules, but they are limited to relatively small systems due to unfavourable scaling of self-energy diagrams with respect to the basis size. We propose an efficient computation of the self-energy diagrams by using tensor-contraction operations to transform the internal summations into functions of external low-level linear algebra libraries. We discuss the achieved computational speed-up in transient electron dynamics in selected molecular systems.
\end{abstract}


\maketitle 


\section{Introduction}\label{sec:intro}

A state-of-the-art computational method for out-of-equilibrium many-body physics is the nonequilibrium Green's function (NEGF) approach~\cite{Baym1961,kbbook,Keldysh1964,Danielewicz1984,svlbook,Balzer2013book}. Mostly due to lack of computational capabilities, the non-linear integro-differential Kadanoff-Baym equations (KBE) for the NEGF from the 1960s remained fairly elusive until their first numerical solutions were presented in 1984 by~\citet{Danielewicz1984b} and further numerical implementations at the turn of the century~\cite{Kohler1999,Semkat1999,Kwong2000}. During the past twenty years a considerable amount of progress has been achieved in various fields of physics employing the NEGF approach: From sub-atomic nuclear reactions~\cite{Dickhoff2004,Rios2011} to atomic and molecular scale~\cite{Dahlen2005,Dahlen2007,Galperin2007,Thygesen2008,Balzer2010a,Balzer2010b,Phillips2014,Perfetto2015,Covito2018,Hopjan2018}, further to condensed phase~\cite{Freericks2009,moritz_electron-mediated_2013,kemper_mapping_2013,Sentef2013,aoki_nonequilibrium_2014,kemper_effect_2014,kemper_direct_2015,sentef_theory_2016,Golez2016,kemper_review_2017,Tuovinen2018} and mesoscopic systems~\cite{Brandbyge2002,Ludovico2012,Wang2014,Ridley2017,Tuovinen2019a,Tuovinen2019b}, and even to descriptions of high-energy particle physics in cosmology~\cite{Kainulainen2002,Garny2009,Garny2013}.

However, combining the KBE with \emph{ab initio} descriptions of realistic materials still remains a computational challenge. This challenge results from the double-time structure of the KBE rendering the method very expensive for both CPU time and storing the objects in RAM. The Generalized Kadanoff--Baym Ansatz (GKBA) offers a simplification by reducing the two-time-propagation of the Green's function to the time-propagation of a time-local density matrix~\cite{Lipavsky1986}. The computational complexity of the time-propagation of the GKBA equations  scales as the number of time steps squared instead of the cubic scaling in the double-time KBE~\cite{Hermanns2012}. When a simulation to reach longer time scales is desired, this difference in computational speed becomes immense. However, this speed-up in computational scaling is only possible for the correlation self-energy approximation at the second-Born (2B) level. The 2B approximation goes beyond mean-field description at the Hartree-Fock (HF) level but it includes the bare interaction only up to second order, i.e., higher order correlations and screening effects are neglected, like in higher order $T$-matrix or $GW$ approximations~\cite{Galitskii1958,Hedin1965}. However, the viability of the 2B approximation has been assessed for a large set of systems with up to moderate interaction strength~\cite{vonFriesen2009,Rusakov2016}.

Even though the above implementations of the NEGF method have been successfully applied in many contexts, the computation of the self-energy still remains a numerical bottleneck. For larger systems to be studied, the scaling with respect to the basis size in the self-energy diagrams may be very unfavourable, making first principles simulations numerically expensive, at least in na{\"i}ve implementations when looping over the full basis. Recently, a dissection algorithm has been proposed and implemented~\cite{Perfetto2019,Perfetto2018} for identifying and utilizing the sparsity of many-body interactions. In this paper we propose to transform the summation expressions in the self-energy diagrams using tensor-contraction operations, and to further employ external linear algebra libraries (e.g. low-level \texttt{C} or \texttt{Fortran}) taking into account, e.g., memory availability, communication costs, loop fusion and ordering~\cite{Baumgartner2005,Solomonik2014,Huang2018}. (Here we consider \emph{tensors} simply as multidimensional objects without deeper (differential-)geometric interpretation.) With benchmark simulations in selected molecular systems we present an efficient way to compute the 2B self-energy applicable either in full time-propagation of the KBE or in the numerically less expensive GKBA variant.

\section{Model and method}\label{sec:model}

We consider a finite and quantum-correlated electronic system described by a time-dependent Hamiltonian
\be\label{eq:hamiltonian}
\hat{H}(t) = \sum_{ij} h_{ij}(t) \hat{c}_i^\dagger \hat{c}_j + \frac{1}{2}\sum_{ijkl}v_{ijkl}(t) \hat{c}_i^\dagger \hat{c}_j^\dagger \hat{c}_k \hat{c}_l ,
\ee
where $i,j,k,l$ label a complete set of one-particle states $\{\varphi(\br)\}$, and $\hat{c}^{(\dagger)}$ are the annihilation (creation) operators for electrons from (to) these states. Although we assume, for simplicity, spin-compensated electrons and invariance under spin rotations, the whole consideration could easily be generalized to include also spin degrees of freedom~\cite{Myohanen2008,Myohanen2009,vonFriesen2009,Latini2014,Giesbertz2016}. The objects henceforth described will be diagonal in spin space. The one-body contribution to the Hamiltonian,
\be\label{eq:onebody}
h_{ij}(t) = \int \ud \br \varphi_i^*(\br) h(\br,t) \varphi_j(\br),
\ee
may have an explicit time dependence, describing, e.g., pump-probe spectroscopies or voltage pulses. These would enter in $h(\br,t) = -\frac{1}{2}\nabla^2 + w(\br,t) - \mu$ as external fields $w$. We also introduced the chemical potential $\mu$ and we absorbed it into the equilibrium description of the one-body part of the Hamiltonian. Atomic units, $\hbar = m = e = 1$, are used throughout. The two-body part accounts for interactions between the electrons with the standard two-electron Coulomb integrals
\be\label{eq:coulomb}
v_{ijkl} = \int \ud \br \int \ud \brp \frac{\varphi_i^*(\br)\varphi_j^*(\brp)\varphi_k(\brp)\varphi_l(\br)}{|\br-\brp|}.
\ee
Even though the Coulomb interaction itself is instantaneous, in Eq.~\eqref{eq:hamiltonian} we allow the strength of the two-body part to be time-dependent to describe, e.g., interaction quenches or adiabatic switching. For real-valued basis functions $\varphi$ the Coulomb integrals in Eq.~\eqref{eq:coulomb} follow $8$-point permutation symmetry
\be\label{eq:symmetries}
v_{ijkl} = v_{jilk} = v_{klij} = v_{lkji} = v_{ikjl} = v_{ljki} = v_{kilj} = v_{jlik},
\ee
which can be verified by permuting dummy integration variables and by complex conjugation. The following discussion is not limited to this choice, however, and also complex and spin-dependent basis functions could be used.

\begin{figure}[t]
\center
\includegraphics[width=0.475\textwidth]{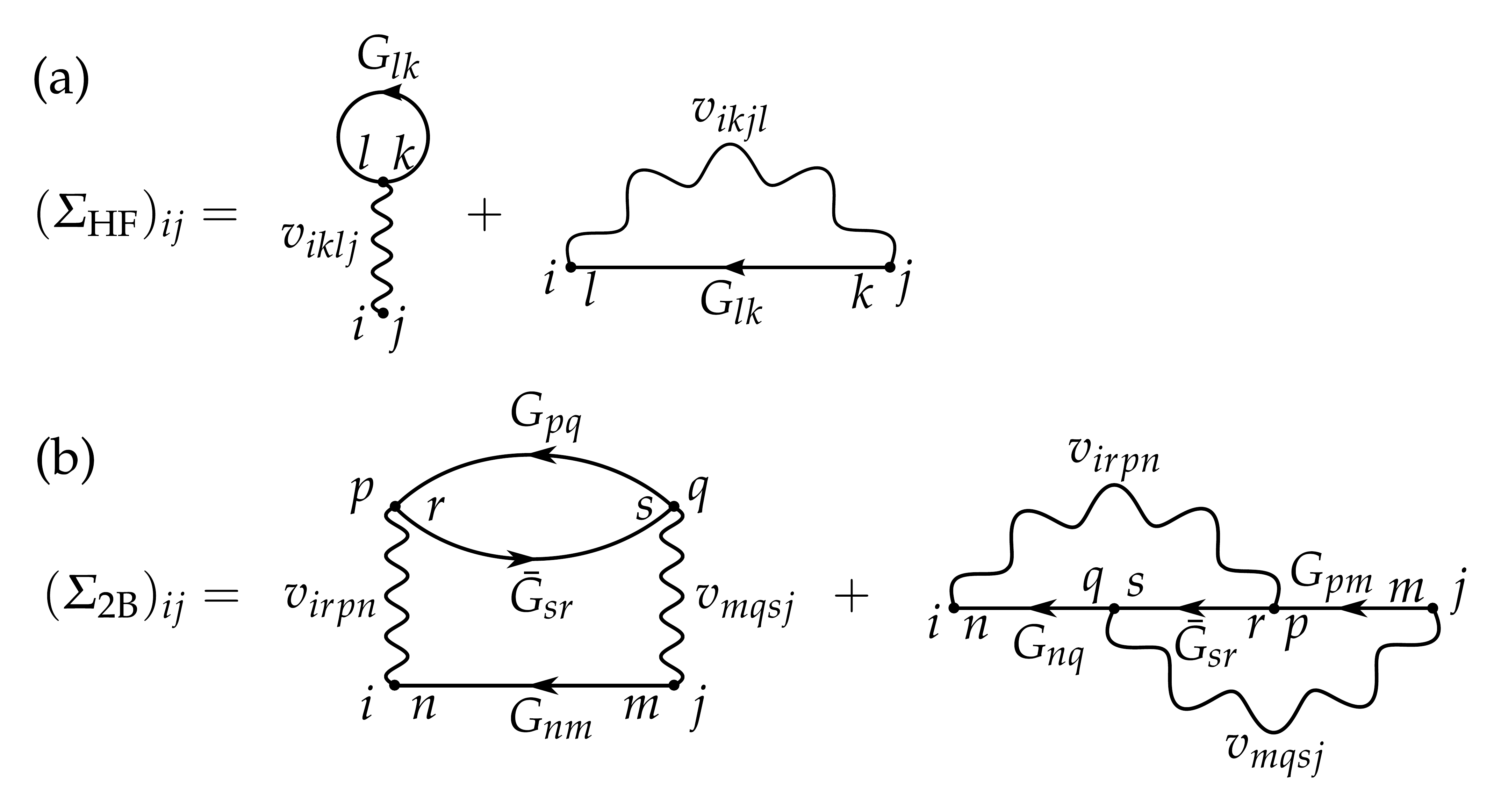}
\caption{Diagrammatic representations of the Hartree-Fock (a) and the second-Born (b) correlation self-energies. The straight lines denote electronic Green's functions whereas the wiggly lines denote the electronic interactions. The internal indices are summed over. Each diagram comes with a prefactor $(-1)^{N_{\text{loop}}}\im^{N_{\text{int}}}$ where $N_{\text{loop}}$ is the number of loops and $N_{\text{int}}$ is the number of interaction lines~\cite{svlbook}. The direct terms with a loop furthermore take an overall spin-degeneracy factor $\xi$, which in this case is $\xi=2$~\cite{Bruus2007,Balzerthesis}.}
\label{fig:diagrams}
\end{figure}

To calculate time-dependent nonequilibrium quantities we use the equations of motion for the one-particle Green's function on the Keldysh contour $\g$~\cite{Danielewicz1984,svlbook,Balzer2013book}. This object is defined as
\be\label{eq:greenf}
G_{ij}(z,z') = -\mathrm{i} \langle {T}_{\gamma} [\hat{c}_i(z) \hat{c}_j^\dagger(z')] \rangle ,
\ee
where $T_\gamma$ is the contour ordering operator and the variables $z,z'$ specify the location of the Heisenberg-picture operators $\hat{c}$ on the Keldysh contour. The contour has a forward and a backward branch on the real-time axis, $[t_0,\infty[$, and also a vertical branch on the imaginary axis, $[t_0,t_0-\im\b]$ with inverse temperature $\b$. The Green's function includes detailed information about particle propagation, and important physical quantities such as electric currents or photoemission spectra can be extracted from it. The Green's function $G$ satisfies the integro-differential equations of motion~\cite{svlbook}
\begin{align}
\left[\im\partial_z - h(z)\right]G(z,z') & = \delta(z,z') + \int_\g \ud \bar{z} \varSigma(z,\bar{z}) G(\bar{z},z') \label{eq:fulleom} \\
G(z,z')\left[-\im\partl_{z'} - h(z')\right] & = \delta(z,z') + \int_\g \ud \bar{z} G(z,\bar{z})\varSigma(\bar{z},z') \label{eq:fulleom2}
\end{align}
where all objects are matrices with respect to the basis of one-particle states $\{\varphi(\br)\}$. The self-energy $\varSigma$ accounts for the electronic interactions. While some two-particle quantities, such as interaction energies and double occupancies, can also be computed from this picture~\cite{Balzerthesis,Balzer2016}, the introduction of the self-energy transforms the many-body problem to an effective quasiparticle picture, and higher order correlations, such as the pair distribution function, are not directly accessible~\cite{vonFriesen2011,Bonitz2013}. Depending on the arguments $z,z'$, the Green's function, $G(z,z')$, and the self-energy, $\varSigma(z,z')$, defined on the time contour have components lesser ($<$), greater ($>$), retarded (R), advanced (A), left ($\lceil$), right ($\rceil$) and Matsubara (M)~\cite{svlbook}. Typically, one concentrates on the particle and hole propagation in terms of $G^<(t,t')$ and $G^>(t,t')$ where the time arguments $t$ and $t'$ refer to the (real) times when a particle is added or removed from the system. Furthermore, the one-particle reduced density matrix (1RDM) is $\rho(t) \equiv -\im G^<(t,t)$ from which one could compute the expectation value of any one-body operator. Taking the equal-time limit ($t'\to t^+$) one obtains from Eqs.~\eqref{eq:fulleom} and~\eqref{eq:fulleom2}
\be\label{eq:equalt}
\im\frac{\ud}{\ud t}G^<(t,t) = [h(t) + \varSigma_{\text{HF}}(t),G^<(t,t)] + I(t) ,
\ee
where we defined the collision integral
\begin{align}\label{eq:collint}
I(t) = \int_{t_0}^t & \ud \bar{t} \left[\varSigma_{\text{c}}^>(t,\bar{t})G^<(\tb,t) - \varSigma_{\text{c}}^<(t,\tb)G^>(\tb,t) \right.\nonumber\\
& \left. + \ G^<(t,\tb)\varSigma_{\text{c}}^>(\tb,t) - G^>(t,\tb)\varSigma_{\text{c}}^<(\tb,t)\right] .
\end{align}
In addition, in Eq.~\eqref{eq:equalt} we separated the time-local and time-non-local contributions to the self-energy as $\varSigma = \varSigma_{\mathrm{HF}} + \varSigma_{\mathrm{c}}$, the former being referred to as the Hartree-Fock (HF) self-energy and the latter the correlation self-energy, see Fig.~\ref{fig:diagrams}. This allows for the extraction of a time-local effective single-particle Hamiltonian, $h(t) + \varSigma_{\text{HF}}(t)$. The collision integrals therefore incorporate only the correlation self-energies $\varSigma_{\text{c}}$. Importantly, the self-energies depend on the Green's functions themselves, $\varSigma[G]$, and therefore the equation of motion needs to be solved self-consistently. The correlation self-energies are typically obtained by a diagrammatic expansion where terms can be systematically summed up to infinite order. In this work we concentrate on the second-Born self-energy, $\varSigma_{\text{c}}=\varSigma_{\text{2B}}$, see Fig.~\ref{fig:diagrams}, but the consideration can be extended to other (higher order) diagrams as well. 

Although we reduced the considered information to the description of a single-time object $\rho$, the double-time nature of the full equations of motion is still present in the collision integral which requires the double-time history of $\varSigma^\lessgtr$ and $G^\lessgtr$ to be stored. In order to obtain a closed equation for $\rho$ it is customary to use the GKBA~\cite{Lipavsky1986}
\be\label{eq:gkba}
G^\lessgtr(t,t') \approx \im\left[G^{\text{R}}(t,t')G^\lessgtr(t',t') - G^\lessgtr(t,t)G^{\text{A}}(t,t')\right] ,
\ee
and an approximation to the double-time propagators $G^{\text{R}/\text{A}}$ at the HF level~\cite{Balzer2013book}
\begin{align}\label{eq:prop}
G^{\text{R}/\text{A}}(t,t') & \approx \mp \im \theta[\pm(t-t')]{T}\ex^{-\im\int_{t_0}^t\ud \tb [h(\tb) + \varSigma_{\text{HF}}(\tb)]} ,
\end{align}
where $T$ is the chronological time-ordering operator~\cite{svlbook}. The HF self-energy, being time-local, can be evaluated from the 1RDM as (see Fig.~\ref{fig:diagrams})
\be
(\varSigma_{\text{HF}})_{ij}(t) = \sum_{kl}(2v_{iklj}-v_{ikjl})\rho_{lk}(t).
\ee
The lesser Green's function or the 1RDM can then be solved from Eq.~\eqref{eq:equalt} by a numerical time-stepping algorithm and using the symmetry property $G^>(t,t) = -\im + G^<(t,t)$~\cite{Hermanns2012,Latini2014,Tuovinen2018}.

In principle, the collision integral on the vertical branch of the Keldysh contour, $I^{\mathrm{ic}}(t) = -\mathrm{i} \int_0^\beta \mathrm{d} \tau \varSigma_{\mathrm{c}}^\rceil (t,\tau) G^\lceil(\tau,t)$, should also be taken into consideration. However, using the GKBA, the initial correlations collision integral, $I^{\mathrm{ic}}$, is usually neglected due to the lack of a GKBA-like expression for the mixed components $G^{\rceil,\lceil}$ and $\varSigma_{\mathrm{c}}^{\rceil,\lceil}$. The correlated initial state therefore needs to be prepared by starting with an uncorrelated (or HF) system and slowly switching on the interaction (adiabatic switching procedure)~\cite{Hermanns2012,Latini2014,Hermanns2014,Tuovinen2018}. However, the inclusion of the initial correlations has been shown to be possible also within GKBA~\cite{Semkat2003,Karlsson2018,Hopjan2019}.

\section{Second-Born self-energy}\label{sec:2b}
For the time-propagation of Eq.~\eqref{eq:equalt} we are only concerned with the lesser and greater components of the Green's function and self-energy. For the sake of notational simplicity, we then write $G \equiv G^\lessgtr(t,t')$, $\bar{G} \equiv G^\gtrless(t',t)$, and $\varSigma \equiv \varSigma_{\text{c}}^\lessgtr(t,t')$. In the second-Born approximation (2B) the correlation self-energy takes the form~\cite{Latini2014,Perfetto2019} (see Fig.~\ref{fig:diagrams})
\begin{align}\label{eq:2bse}
\varSigma_{ij} & = 2 \sum_{\substack{mn\\pq\\rs}} v_{irpn} v_{mqsj} G_{nm} \bar{G}_{sr} G_{pq} \nonumber \\
& - \sum_{\substack{mn\\pq\\rs}} v_{irpn} v_{mqsj} G_{nq} \bar{G}_{sr} G_{pm}.
\end{align}
As can be seen from Eq.~\eqref{eq:2bse} computing the full self-energy matrix by direct looping takes $N_b^8$ operations where $N_b$ is the size of the basis. However, it is possible to reduce this scaling to $\propto N_b^5$ by grouping and reorganizing the objects in Eq.~\eqref{eq:2bse}~\cite{Yang2011,Hermannsthesis,Neuhauser2017,Karlsson2018,Perfetto2019,Schluenzen2019}. We address this more thoroughly in Sec.~\ref{sec:subsec-gen}. It is also to be noted that the 2B self-energy is non-local in time, i.e., this computation needs to be performed for two times $t$ and $t'$, and it is important to keep track of the correct time arguments in the objects $v$, $G$ and $\bar{G}$. While the 2B approximation together with the GKBA allows for a maximal speed-up in computational scaling compared to the full two-time KBE ($T^2$ vs. $T^3$, $T$ being the total propagation time), GKBA simulations with $GW$ and $T$-matrix approximations to the self-energy have also been performed~\cite{Schluenzen2016,Schluenzen2017}.

We note that the NEGF method using the 2B self-energy, sometimes referred to as second-order Green's function (GF2)~\cite{Holleboom1990,Phillips2014,Rusakov2016,Neuhauser2017}, can be related to the second-order M{\o}ller-Plesset perturbative expansion (MP2)~\cite{Helgaker2000}. The construction of the MP2 correction is similar to the 2B self-energy, although in the NEGF approach the self-energy enters nonlinearly into an updated Green's function and this procedure is continued until convergence is reached, whereas the MP2 can be related to the first step of this iteration~\cite{Holleboom1990,Phillips2015,Neuhauser2017}.

Next, we consider three different cases for the interaction vertex: (1) Diagonal basis where the Coulomb integrals take the Hubbard-like form $v_{ijkl} = U_{i}\delta_{ij}\delta_{ik}\delta_{il}$; (2) Symmetric basis where the Coulomb integrals allow for non-diagonal or long-range interactions $v_{ijkl} = V_{ij}\delta_{il}\delta_{jk}$ but the $4$-point vertex is symmetric (density-density type interaction); and (3) The general basis of the full Coulomb integral $v_{ijkl}$. From the resulting structures of the internal summations in the self-energy diagrams, we identify matrix or tensor operations. Instead of simply looping over the basis indices, employing well-established linear algebra libraries for the matrix and tensor operations~\cite{Baumgartner2005,Solomonik2014,Huang2018} may speed up the construction of the self-energy.

We denote matrix multiplication by ``$\times$'' and entrywise multiplication (Hadamard or Schur product) by ``$\circ$''. For example, in \texttt{Fortran} and \texttt{Mathematica} the entrywise products are done through simple multiplication operator ``$*$'' whereas the matrix product is done through the ``\texttt{matmul}'' or ``$.$'' operators. In \texttt{C++} with the \texttt{Armadillo} library~\cite{armadillo} the symbol ``\%'' is used for entrywise products whereas ``$*$'' is a matrix product. In \texttt{Python} with the \texttt{NumPy} (\texttt{np}) numerical library~\cite{numpy} the entrywise product can be done with the function ``\texttt{np.multiply}'' whereas the matrix or more general tensor multiplication can be done via the ``\texttt{np.dot}'' or the ``\texttt{np.einsum}'' functions.

\subsection{Diagonal basis}
For a diagonal basis, $v_{ijkl} = U_{i}\delta_{ij}\delta_{ik}\delta_{il}$, Eq.~\eqref{eq:2bse} is simplified as
\be\label{eq:2b-diag}
\varSigma_{ij} = U_i U_j G_{ij} \bar{G}_{ji} G_{ij},
\ee
and the computational cost of constructing the full matrix therefore scales as $N_b^2$. In this simple case there are no further contractions to perform as the internal summations were already explicitly resolved due to the Kronecker $\delta$'s in the interaction vertex. Because in many practical implementations entrywise multiplication between two objects is only possible when they have the same dimension, we rewrite the on-site interaction $U_i$ instead as the diagonal part $V_{ii}$ of a matrix. The resulting expression can then be recasted in matrix form as an entrywise product
\be\label{eq:2b-diag-opt}
\varSigma = \mathrm{diag} (V) \circ \mathrm{diag} (V) \circ G \circ \bar{G}^{\text{T}} \circ G .
\ee
We anticipate that this is a faster construction for the whole self-energy matrix instead of looping over the basis indices $i,j$ in Eq.~\eqref{eq:2b-diag} when passing the matrix operations in Eq.~\eqref{eq:2b-diag-opt} to an external linear algebra library.

\subsection{Symmetric basis}
For a symmetric basis, $v_{ijkl} = V_{ij}\delta_{il}\delta_{jk}$, Eq.~\eqref{eq:2bse} is simplified as
\be\label{eq:2b-symm}
\varSigma_{ij} = 2\sum_{kl} V_{i k} V_{j l} G_{i j} \bar{G}_{l k}  G_{k l} - \sum_{kl} V_{i k} V_{j l} G_{i l} \bar{G}_{l k} G_{k j} .
\ee
We first consider the first term of Eq.~\eqref{eq:2b-symm}, i.e., the second-order bubble diagram, and visualize the contraction path for efficient computation. The expression can be manipulated as
\begin{align}
\varSigma^{\text{b}}_{ij} & = 2\sum_{kl} V_{i k} V_{j l} G_{i j} (\bar{G}^{\text{T}})_{k l}  G_{k l} \nonumber \\
& = 2\sum_{kl} V_{i k} V_{j l} G_{i j} (\bar{G}^{\text{T}} \circ G)_{k l} \nonumber \\
& = 2\sum_{l} V_{j l} G_{i j} \sum_k V_{i k} (\bar{G}^{\text{T}} \circ G)_{k l} \nonumber \\
& = 2\sum_{l} V_{j l} G_{i j} [V \times (\bar{G}^{\text{T}} \circ G)]_{il} \nonumber \\
& = 2\sum_{l} V_{j l} G_{i j} \{[V \times (\bar{G}^{\text{T}} \circ G)]^{\text{T}}\}_{li} \nonumber \\
& = 2G_{i j} \sum_{l} V_{j l} \{[V \times (\bar{G}^{\text{T}} \circ G)]^{\text{T}}\}_{li} \nonumber \\
& = 2G_{i j} (V \times \{[V \times (\bar{G}^{\text{T}} \circ G)]^{\text{T}}\})_{ji} \nonumber \\
& = 2G_{i j} [(V \times \{[V \times (\bar{G}^{\text{T}} \circ G)]^{\text{T}}\})^{\text{T}}]_{ij} \nonumber \\
& = 2\{G \circ [(V \times \{[V \times (\bar{G}^{\text{T}} \circ G)]^{\text{T}}\})^{\text{T}}]\}_{ij},\label{eq:bubble-contract}
\end{align}
where we identified matrix transposes, entrywise products and matrix multiplications. The procedure outlined above, unfortunately, makes the final expressions less readable, but in the end the full self-energy matrix (for the bubble diagram part) may be constructed as a one-liner $\varSigma^{\text{b}} = 2G \circ [(V \times \{[V \times (\bar{G}^{\text{T}} \circ G)]^{\text{T}}\})^{\text{T}}]$. However, as mentioned earlier, one must keep track of the time arguments, i.e., reading from left the first $V$ is evaluated at $t'$ and the second $V$ is evaluated at $t$.

Contractions on the internal summations in the self-energy diagrams do not always yield a favourable path. If we take the second term in Eq.~\eqref{eq:2b-symm}, i.e., the second-order exchange diagram, obtaining an expression similar to Eq.~\eqref{eq:bubble-contract} is not possible for the full self-eneregy matrix. However, for the diagonal part of the exchange diagram we obtain 
\begin{align}
\varSigma^{\text{x}}_{ii} & = -\sum_{kl} V_{i k} V_{i l} G_{i l} \bar{G}_{l k} G_{k i} \nonumber \\
& = -\sum_{l} (V \circ G)_{i l} \sum_k \bar{G}_{l k} (V^{\text{T}} \circ G)_{k i} \nonumber \\
& = -\sum_{l} (V \circ G)_{i l} [\bar{G} \times (V^{\text{T}} \circ G)]_{l i} \nonumber \\
& = -\{(V \circ G) \times [\bar{G} \times (V^{\text{T}} \circ G)]\}_{i i}.\label{eq:x-contract}
\end{align}
The off-diagonal parts would still need to be evaluated by explicit looping as in Eq.~\eqref{eq:2b-symm}, but the above contraction path may also be combined with, e.g., the dissection algorithm of Ref.~\cite{Perfetto2019} where chosen pairs of the Coulomb integral matrix elements (according to some cut-off energy) would be used. This further reduces the requirement for looping over the basis indices.

\subsection{General basis}\label{sec:subsec-gen}
For a general basis all $v_{ijkl}$ are nonvanishing. In this case the multi-index summations in the self-energy diagrams and their consequent contractions are not always easy to see, but this task can be automatized using, e.g., the \texttt{np.einsum\_path} function in \texttt{Python}. The information obtained for an optimal sequence of contractions may further be combined with the symmetry properties~\eqref{eq:symmetries} and with a pre-determined subset of nonzero Coulomb integrals~\cite{Perfetto2019}. 

Manipulating Eq.~\eqref{eq:2bse} gives
\begin{align}\label{eq:2b-opt}
\varSigma_{ij} & = 2\sum_{\substack{np\\qs}} G_{pq} \underbrace{\sum_m v_{mqsj}G_{nm}}_{\equiv T_{nqsj}^{(1)}} \ \underbrace{\sum_{r} v_{irpn}\bar{G}_{sr}}_{\equiv T_{ispn}^{(2)}} \nonumber \\
& - \sum_{\substack{mn\\ps}} G_{pm} \underbrace{\sum_{q} \overbrace{v_{mqsj}}^{=v_{qmjs}} G_{nq}}_{= T_{nmjs}^{(1)}} \underbrace{\sum_{r}v_{irpn}\bar{G}_{sr}}_{= T_{ispn}^{(2)}} \nonumber \\
& = 2\sum_{nqs} T_{nqsj}^{(1)} \underbrace{\sum_p G_{pq} T_{ispn}^{(2)}}_{\equiv T_{isqn}^{(3)}} - \sum_{mns} T_{nmjs}^{(1)} \underbrace{\sum_p G_{pm}T_{ispn}^{(2)}}_{= T_{ismn}^{(3)}} \nonumber \\
& = \sum_{nqs} (2T_{nqsj}^{(1)} T_{isqn}^{(3)} - T_{qnjs}^{(1)}T_{isnq}^{(3)}) ,
\end{align}
where we defined tensor contractions $T^{(1,2,3)}$ and permuted indices with the help of Eq.~\eqref{eq:symmetries}, identifying similar contractions consequently. We see from the last line of Eq.~\eqref{eq:2b-opt} that for constructing the full self-energy matrix the scaling over the basis is reduced from $N_b^8$ to $\propto N_b^5$~\cite{Hermannsthesis,Neuhauser2017,Perfetto2019,Schluenzen2019}.

\begin{figure}[t]
\center
\includegraphics[width=0.45\textwidth]{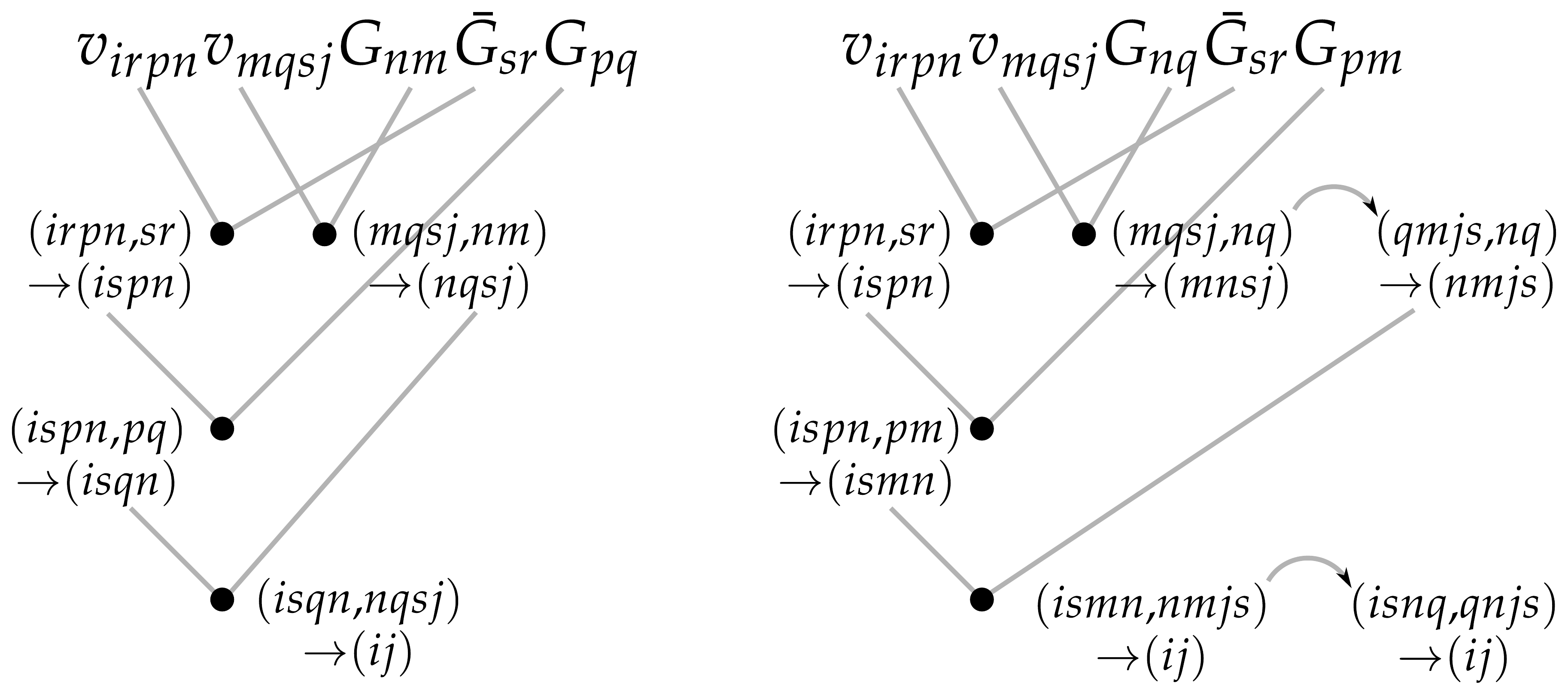}
\caption{Contraction paths for the computation of the self-energy in Eq.~\eqref{eq:2b-opt}. The dots denote tensor-contraction operations which could be implemented, e.g., using the \texttt{np.einsum} function in \texttt{Python} which includes (from version $1{.}14$ onwards) optimized ordering and dispatching many operations to canonical \texttt{BLAS} routines\cite{opteinsum}.}
\label{fig:contractions}
\end{figure}

As before, the readability of the self-energy in Eq.~\ref{eq:2b-opt} suffers a bit compared to Fig.~\ref{fig:diagrams} or Eq.~\eqref{eq:2bse}. However, Eq.~\eqref{eq:2b-opt} is visualized in Fig.~\ref{fig:contractions}, and for the sake of efficient computation the contraction operations can be grouped together and executed essentially as a single command, where the lower-level loop fusions and orderings of operations are handled by the underlying numerical library. 
We emphasize that while the reorganizations of the summations in Eq.~\eqref{eq:2bse} to arrive at Eq.~\eqref{eq:2b-opt} have already been considered to some degree in Refs.~\cite{Hermannsthesis,Perfetto2019,Schluenzen2019}, here we concentrate on the practical computation of the self-energy by employing efficient tensor-contraction operations with a possible contraction path shown in Fig.~\ref{fig:contractions}. Alternative contraction paths than the one shown in Fig.~\ref{fig:contractions} are also possible.

\section{Numerical benchmarks}\label{sec:num}
For the three different cases presented in the previous section, (1) diagonal, (2) symmetric and (3) general bases, we now present sample numerical simulations for the purpose of benchmarking and assessing the validity and accuracy of the alternative implementations of the 2B self-energy. For test cases we choose molecular systems falling into each of the categories: $1$D Hubbard chains which can be related to, e.g., conjugated polymers~\cite{Kirtman1997,vanFaassen2002,vanFaassen2003,Hermanns2014} with local (1) and long-range interactions (2). We set the hopping energy between nearest-neighbors $J=-1$, the on-site electron-electron interaction $U=1$, and the long-range interaction between particles at atomic sites $i$ and $j$ as in the Ohno model $V_{ij} = U/\sqrt{1+|i-j|^2}$~\cite{Ohno1964,polymer}. For the case (3) we take a $\text{CH}_4$ molecule with a general one-particle Kohn-Sham (KS) basis obtained from density-functional theory (DFT) using \texttt{Octopus}~\cite{Marques2003}. Using this DFT calculation, the one- and two-body matrix elements [Eqs.~\eqref{eq:onebody} and~\eqref{eq:coulomb}] are then constructed in the corresponding KS basis; a more detailed explanation can be found in Ref.~\cite{Perfetto2018}.

We implement the explicit loops over the basis indices [Eqs.~\eqref{eq:2b-diag},~\eqref{eq:2b-symm}, and~\eqref{eq:2b-opt}] in \texttt{C++}. In the cases (1) and (2) we employ the matrix operations [Eqs.~\eqref{eq:2b-diag-opt},~\eqref{eq:bubble-contract}, and~\eqref{eq:x-contract}] using the \texttt{Armadillo} library (version 9.200.5)~\cite{armadillo}, and in the case (3) we employ the tensor operations [Eq.~\eqref{eq:2b-opt} and~Fig.~\ref{fig:contractions}] using the \texttt{NumPy} library (version 1.15.1) in \texttt{Python}~\cite{numpy}. We perform the comparisons using a regular desktop computer with an Intel Core i5-4460 @ 3.2 GHz with 6 MB cache, running on 64-bit architecture using Ubuntu 18.04 operating system incorporating the Linux kernel 4.15.0 and the GCC 7.3.0 compiler. The comparisons are done using only a single core to better benchmark the computational cost.

\begin{figure*}[t!]
\center
\includegraphics[width=0.72\textwidth]{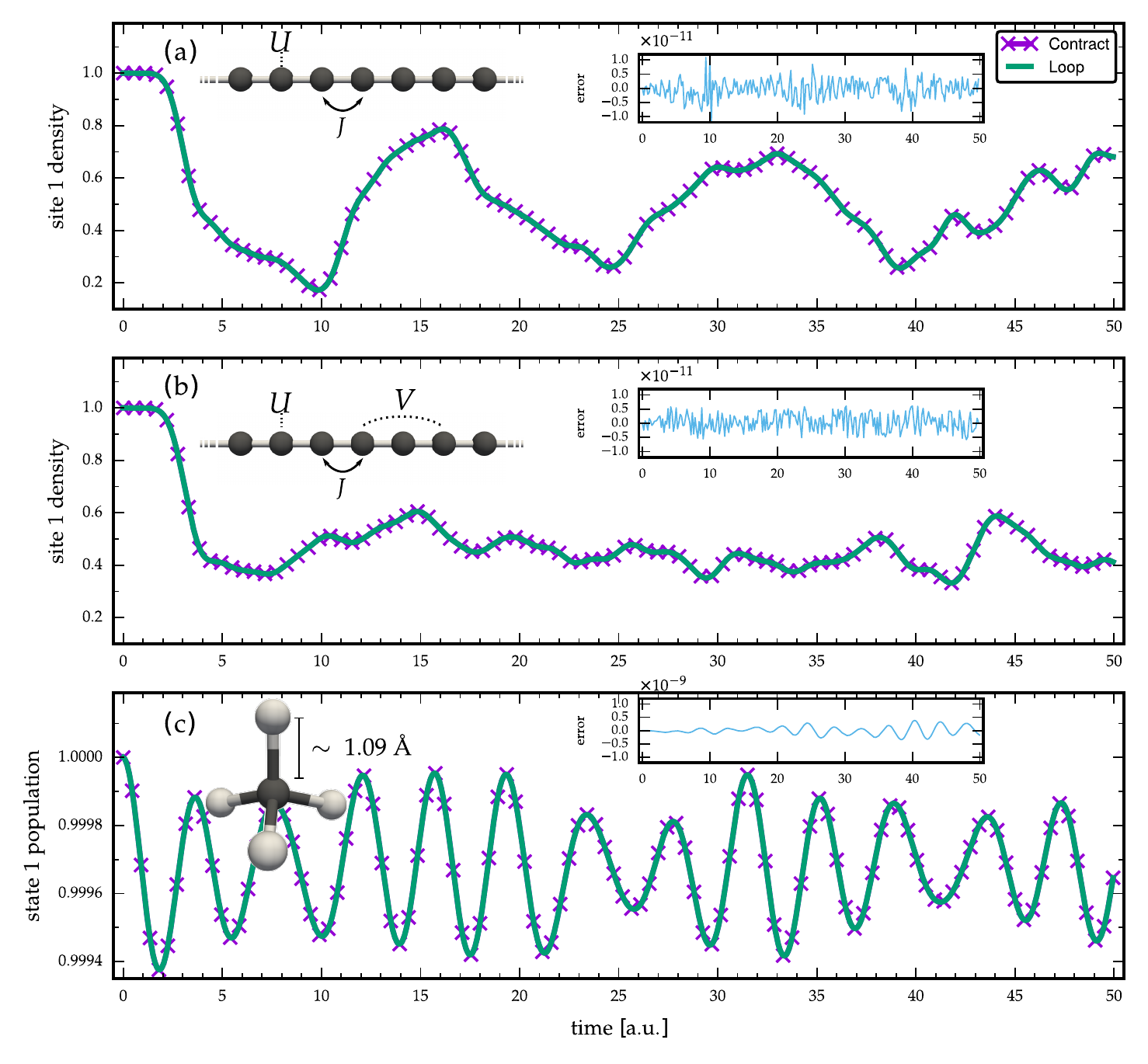}
\caption{Time-dependent 1RDM elements for the three different systems studied: (a) Diagonal basis with local interaction, (b) Symmetric basis with local and long-range interaction, and (c) General basis with the full Coulomb integrals. The insets show the relative difference between the two curves in the main plots.}
\label{fig:simulations}
\end{figure*}

We perform a time-propagation {\`a} la GKBA of $N_t$ time steps with length $\delta$. For the sake of simpler computation, in this work we do not employ any predictor-corrector schemes. For the polymer chain we take $N_b=10$ atomic sites and start the time-propagation from an initial state where $N_b/2$ particles are trapped to the $N_b/2$ leftmost sites by applying a strong confinement potential~\cite{Hermanns2014}. This configuration relaxes once the time evolution is started. For the $\text{CH}_4$ molecule we represent the $4$ electrons by $N_b=10$ basis functions, and we start the time-propagation from a HF initial state, which can be obtained from a separate (time-independent) calculation, and then suddenly switch on the many-body correlations in the 2B self-energy. This sudden process can be interpreted as an interaction quench introducing transient dynamics. 

For the case (1) we take $N_t=5000$ time steps of length $\delta=0.01$, for the case (2) $N_t=2000$ time steps of length $\delta=0.025$, and for the case (3) we take $N_t=1000$ time steps of length $\delta=0.05$. The reason for the varying number of time steps between the investigated cases is that a calculation with $N_t=1000$ would be too fast to execute in case (1) for a meaningful comparison of runtimes, whereas $N_t>1000$ in case (3) would lead to unpractically long execution times for the sake of the present study. Here we are not too concerned about the physical mechanisms taking place during the transient oscillations or how accurate the 2B self-energy is compared to more sophisticated approximations, but our aim is simply to assess the validity of the proposed computation scheme, and to compare execution runtimes.

In Fig.~\ref{fig:simulations} we show the transient dynamics of the three cases discussed above. The execution runtimes for each of these simulations are shown in Tab.~\ref{tab:runtimes}. We confirm that within numerical accuracy, both looping over the basis indices and employing tensor-contraction operations, give the same result. Importantly, the execution runtimes are brought down by employing the tensor-contraction operations in the computation of the 2B self-energy. Furthermore, we have checked by increasing the number of time steps that the runtimes increase accordingly, i.e., the gain factors in Tab.~\ref{tab:runtimes} remain roughly similar. For additional validation we have compared our data in Fig.~\ref{fig:simulations}(c) against the \texttt{CHEERS} code~\cite{Perfetto2018} and we find perfect agreement. We note in passing that an ill-advised looping over the full basis in Eq.~\eqref{eq:2bse} ($\propto N_b^8$) instead of the reduced looping in Eq.~\eqref{eq:2b-opt} ($\propto N_b^5$) would result in considerably higher execution runtimes.

\begin{table}
\begin{tabular}{c|c|c|c|c}
Basis & \ Scaling \ & \ Time (loop) \ & \ Time (contr.) \ & \ Gain \\
\hline
\hline
diagonal & $N_b^2$ & 177 & 164 & 1{.}08 \\
symmetric \ & $N_b^4$ & 1213 & 731 & 1{.}66 \\
general & $N_b^5$ & 1527 & 1333 & 1{.}15
\end{tabular}
\caption{Comparison of serial runtimes (in seconds) of sample simulations of basis size $N_b=10$ when calculating the self-energy by looping over the basis indices or employing tensor-contraction operations. The gain factor is defined as the ratio of the runtimes. (Note that different number of time steps is taken for the different lines for better comparison of the runtimes.)}
\label{tab:runtimes}
\end{table}

\begin{figure}[h!]
\centering
\includegraphics[width=0.35\textwidth]{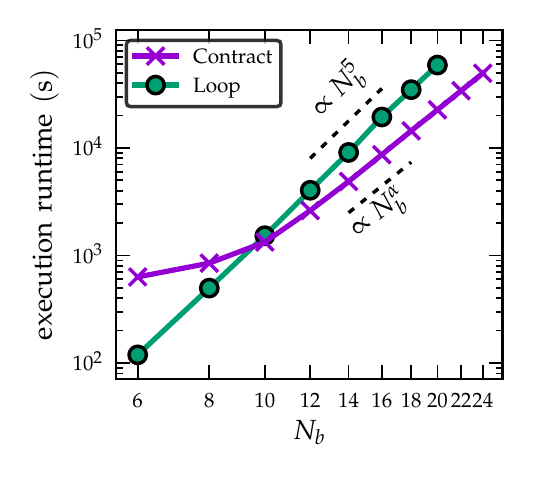}
\caption{Runtime scaling corresponding to Fig.~\ref{fig:simulations}(c) with increasing number of basis functions $N_b$. Using tensor-contraction operations we find a reduced scaling law $\overset{\sim}{\propto} N_b^{4.3}$ compared to the explicit looping over the basis ($\propto N_b^5$).}
\label{fig:scaling}
\end{figure}

As the number of basis functions $N_b=10$ was relatively small in the previous calculations, we expect increase in the gain factors when larger basis is used, due to profiting more from the optimized underlying numerical libraries. In Fig.~\ref{fig:scaling} we show the execution runtimes corresponding to Fig.~\ref{fig:simulations}(c) but with varying number of basis functions. With explicit looping over the basis indices we observe $\propto N_b^5$ behaviour. For smaller basis sizes the explicit looping is faster compared to the tensor-contraction operations done on the \texttt{NumPy} arrays. However, for larger basis sizes the runtimes using the tensor-contraction operations are significantly smaller, also following a power law behaviour $\propto N_b^\a$ for which we empirically find $\a\approx 4.34 \pm 0.17$, see Fig.~\ref{fig:scaling}. This exponent and its statistical errors were extracted by performing a nonlinear least squares fit to the flat part, $N_b\in[14,24]$, using \texttt{gnuplot}. This reduced scaling could be related to the optimization of matrix multiplication using the Strassen algorithm~\cite{Strassen1969}, and to more advanced methods for tensor contraction algorithms which can scale faster than the na{\"i}ve looping scheme~\cite{Huang2018}.

\section{Conclusion}\label{sec:concl}

We presented an efficient way to compute the 2B self-energy diagrams, in the NEGF approach, by using tensor-contraction operations. The apparent attraction for efficient computation of the 2B self-energy, in particular, was due to the maximal speed-up in computational scaling when used together with the GKBA. The internal summations in the self-energy calculations were transformed into matrix and tensor operations to be performed by external low-level linear algebra libraries, speeding up the computation. We anticipate the speedup may be even more advantageous when the code is executed in parallel, taking full advantage of the optimized underlying numerical libraries. Instead of looping over the basis indices, utilizing efficiently optimized external numerical libraries for the tensor-contraction operations has the further advantage of speeding up the computation if/when future implementations of the the external libraries become faster and even more efficient~\cite{Huang2018}.

There has been recent progress in reducing the computational bottleneck of constructing various self-energy approximations by using stochastic methods~\cite{Ge2014,Neuhauser2014,Neuhauser2017}. Here we mention the work of~\citet{Neuhauser2017} who considered the 2B self-energy in an equilibrium setting and achieved a much more favourable quadratic scaling over the fifth power. While the reduced scaling with respect to the basis size using these stochastic methods goes beyond our findings, it is not straightforward to argue how the accuracy of such a stochastic-sampling approach may affect convergence or error propagation in an out-of-equilibrium setting. In this case one would have to sample not a single $\tau$-axis (Matsubara) self-energy but instead a new slice of ever-expanding self-energies $\varSigma_c^\lessgtr(t,t')$ in the two-time plane. However, it would be a promising venue to extend the stochastic methods also to real time in future studies~\cite{Ruan2018}.

The presented approach is not limited to the 2B self-energy only but could be readily used for other correlation self-energies, such as $GW$ or $T$-matrix. In addition, many other similar multi-index operations, such as evaluating the initial correlations collision integral in Ref.~\cite{Karlsson2018}, might become computationally more accessible by using the tensor-contraction representations. In the present work we considered only the GKBA with Hartree-Fock propagators, but extensions to correlated approximations to the propagator~\cite{Latini2014} are also directly applicable in our approach. The presented simulations in selected molecular systems provided concrete evidence of the accuracy and applicability of the tensor-contraction operations. With reasonable and precise implementations or variations of the present study, we expect this procedure to allow for considerably larger basis sizes to be possible to address in forthcoming NEGF+first principles simulations.

\begin{acknowledgments}
R.T. and M.A.S. acknowledge funding by the DFG (Grant No. SE 2558/2-1) through the Emmy Noether program. We wish to thank Damian Hofmann, Gianluca Stefanucci, Michael Bonitz, and Angel Rubio for productive discussions.
\end{acknowledgments}


%

\end{document}